\documentclass[10pt,a4paper]{article}

\title{Edge colouring Game on Trees with maximum degree $\Delta=4$}

\author{Akshay Singh\thanks{Presently with Mastercard, India}
{\normalsize{and }} Sanjeev Saxena\thanks{E-mail:
ssax@cse.iitk.ac.in}\\ Dept. of Computer Science and Engineering,\\
Indian Institute of Technology,\\ Kanpur, INDIA-208 016}

\date{} 

\begin{document}
\maketitle

\subsection*{\centering{Abstract}}

Consider the following game. We are given a tree $T$ and two players
(say) Alice and Bob who alternately colour an edge of a tree (using
one of $k$ colours). If all edges of the tree get coloured, then Alice
wins else Bob wins. Game chromatic index of trees of is the smallest
index $k$ for which there is a winning strategy for Alice. If the
maximum degree of a node in tree is $\Delta$, Erdos et.al.[6], show
that the game chromatic index is at least $\Delta+1$. The bound is
known to be tight for all values of $\Delta\neq 4$.

In this paper we show that for $\Delta=4$, even if Bob is allowed to
skip a move, Alice can always choose an edge to colour and win the
game for $k=\Delta+1$.  Thus the game chromatic index of trees of
maximum degree $4$ is also $5$. Hence, game chromatic index of trees
of maximum degree $\Delta$ is $\Delta+1$ for all $\Delta\geq 2$.

Moreover,the tree can be preprocessed to allow Alice to pick the next
edge to colour in $O(1)$ time.

A result of independent interest is a linear time algorithm for
on-line edge-deletion problem on trees.

\section{Introduction}

We are given a tree or forest in which maximum degree of a node is
$\Delta$. Two players (say) Alice and Bob alternately properly colour
an edge of the forest (using $\Delta+1$ colours)-- a colouring is
proper if no two edges with same colour are incident at a node.  If
all edges get coloured, then Alice wins else Bob wins. 

Bodlaender [3], introduced vertex-colouring game on graphs-- each
player in turn colours a vertex of graph $G$ from $k$ colours.  Alice
wins if the graph $G$ gets completely coloured Bob wins if there is at
least one uncoloured vertex which can not be coloured using $k$
colours. The game chromatic number $\chi_g(G)$ is the smallest number
$k$ such that Alice can always colour $G$ completely. 

Cai and Zhu[4] introduced the corresponding edge-colouring game.  Game
chromatic index $\chi'(G)$ is the minimum number of colours for which
Alice has a winning strategy if the game is played on a graph $G$ [4].
From result of Cai and Zhu[4] if $\Delta=3$ then the game chromatic
index is at most $4$ if there are odd number of edges in the forest.  
Erdos et.al.[6] show that at least $\Delta+1$ colours are required for
a tree of maximum degree $\Delta$, thus for any tree $T$,
$\chi'(T)\geq \Delta+1$. They also show that the bound is tight for
$\Delta\geq 6$; they observe that who moves first is irrelevant for
this game. Andres[1] shows that for $\Delta=5$, the bound of
$\Delta+1$ is also tight. Thus, game chromatic index of trees with
maximum degree $\Delta$ is $\Delta+1$ for $\Delta\neq 4$. For
$\Delta=4$ case, Chan and Nong[5] show that the bound is tight if
either the trees are caterpillar or no two vertices of degree four are
adjacent. Fong et.al.[8] show that the bound is true provided each
degree $4$ node has at most one other neighbour of degree $4$.
Andre[2], Chan and Nong[5] and Fong et.al [8] allow Bob to skip moves.
Fong and Chan [9] study the problem using colours greater than the
game chromatic index. 

We show that for $\Delta=4$ and $\Delta=5$, even if Bob is allowed to
skip a move, Alice can always choose an edge to colour and win the
game. Moreover, the edge can be chosen in $O(1)$ time. Proposed
strategy for Alice is very similar to the technique used by Erdos
et.al [6] for $\Delta\geq 6$ case. 

A result of independent interest is a linear time algorithm for
on-line edge-deletion problem on trees. Even and Shiloach[7] show 
that on-line edge deletion problem on trees can be solved in $O(n \log
n)$ time. We show that the time can be reduced to $O(n)$.

In next section, we summarise the notation used; this is by and large
the notation used by Erdos et.al[6]. In Section 3, we
describe the proposed strategy. In Section 4, we show that the edge to
be coloured can be chosen in $O(1)$ time. We also propose a faster
solution to on-line edge deletion problem on trees in Section 4.

\section{Notation}

We by and large follow the notations of Erdos et.al.[6].

\begin{description}

\item[Leaf edge] An edge which is incident at a leaf node is called
a leaf edge.

\item[{Induced sub-tree}] The induced sub-tree of a component
has 
\begin{itemize}
\item the coloured edges present in the original component and 
\item path of uncoloured edges connecting them.  
\end{itemize}

\item[{Base node}] A node which has degree $3$ or more in the induced
sub-tree of a component is a base node of that component. 

\item[{Star-like}] A component with a unique base node in the induced
sub-tree of that component is called star-like component. If the
star-like component has $x$ coloured edges then it is called as
{\em{$x$-cl edge star}}. 

\item[{Relevant $x$-cl edge star}] A $x$-cl edge star with a unique
base node is relevant if the base node of the $x$-cl edge star has
degree exactly equal to $\Delta$ (maximum degree in the forest) in the
original component. The base node of a relevant x-cl edge star is
called {\em{relevant base node}}. The relevant base node need not have
degree $\Delta$ in the induced sub-tree.

\item[{Matched edge}] In a component $T$ with a unique base node $v$
matched edges are
\begin{itemize}
\item all coloured edges incident on the base node $v$ and

\item other coloured edges whose colour is the same as colour of some edge
incident on the base node $v$
\end{itemize}

\item[{Unmatched edge}] A coloured edge which is not matched will be
called unmatched. These are the edges in $T$ which are not incident on
a base node and whose colour is different from all coloured edges
incident on base node.\\

 We also define

\item[$x$-cl edge component] If the number of coloured edges in a
component (tree) $T$ is $x$ then we call the component $T$ a $x$-cl
edge component ($x$-cl edge tree).

\end{description}

\section{Game Chromatic Index of Forest of degree $4$ or $5$}

If an edge (say) $uv$ of a tree $T$ is coloured, then as in [6], the
tree $T$ is split into two parts. One part (say) $T'$ contains the
vertex $u$ and the other part (say) $T''$ contains the vertex $v$. The
edge $uv$ is then ``added'' to both $T'$ and $T''$.  Basically we
first remove the edge $uv$ from $T$ to split $T$ into two trees $T'$
and $T''$, and then add a copy of this edge to both the trees. Thus
all coloured edges will be leaf edges in some component; internal
edges are always uncoloured.

In rest of this paper we assume that the maximum degree $\Delta$ is
either $4$ or $5$.

We show that after each move of Alice following conditions hold. We
keep the first condition $(S)$ as the same as that in Erdos et.al.[6]
but use a more strict second condition $(M)$:

\begin{description}

\item[Condition {$(S)$}] Every $x$-cl edge component with $x\geq 3$ is
star-like.

\item[Condition {$(M)$}] Every relevant $x$-cl edge star has at most
$\max(3 - \gamma, 0)$ unmatched coloured edges. Here $\gamma $ is the
number of coloured edges incident on the base node (and $x$ is the
total number of coloured edges in the $x$-cl edge star).
\end{description}

Remark 1: Even when $\max(3-\gamma,0)$ is $3$, we ensure that after
Alice's move at most two edges in any component are unmatched.

Remark 2: In each move an uncoloured edge of the original tree (or a
component of the original tree is chosen). However, the conditions are
enforced only in the induced sub-tree. All coloured edges are always
present in some induced sub-tree.

We show by induction on the number of moves that Alice can maintain
the invariants $(M)$ and $(S)$.  Let us assume that invariant $(M)$
and $(S)$ were true after Alice's previous move. We show that they are
also true after the next move of Alice, i.e. after a move of Bob and
then the move by Alice or in case Bob skips his move, then after the
previous move of Alice.  

We consider different cases based on the value of $x$ in $x$-cl edge
components. First in Section 3.1, we consider the case, when there are
at most two coloured edges in each component. We will be left with the
case when some component has at least three coloured edges. In Section
3.2, we consider the case when the two conditions are true before
Alice's move. In Section 3.3 we consider the case when condition $(S)$
is violated and finally in Section 3.4 we finally consider the case
when condition $(S)$ holds but condition $(M)$ gets violated.
 
\subsection{$x$-cl edge components with $x \leq 2$} 

We first consider the case when there are $x$-cl edge components with
uncoloured edges and $x \leq 2$. Then Alice moves as follows: 
\begin{enumerate}

\item If all $x$-cl edge components have no coloured edges ($x=0$),
then Alice colours any edge $uv$. The tree containing edge $uv$
gets split and two $1$-cl edge components are formed. Conditions $(M)$
and $(S)$ remain vacuously satisfied.

\item Assume that an $x$-cl edge component contains exactly one
($x=1$) coloured edge (say) $uv$ and that component is not fully
coloured. If $v$ is a leaf, Alice colours an uncoloured edge incident
at $u$ (else she colours an edge incident at $v$). Thus degree of $u$
in the induced sub-tree becomes two.  As there is no base node
conditions $(M)$ and $(S)$ are vacuously satisfied.

\item Finally consider the case when some $x$-cl edge component
contains two coloured edges $(x=2)$. As there are $2$ coloured edges
(leaf edges), the $x$-cl edge component just contains the path (in the
induced sub-tree) between these two leaf edges; the path may contain
zero or some uncoloured edges. There are two sub-cases:

\begin{enumerate}

\item First consider the case when path has at least one uncoloured
edge. Alice colours an edge on this path which is adjacent to either
coloured edge using a colour different from the colour(s) already
present in the component. 

After colouring the edge, the path gets split in two new $2$-cl edge
component (two paths with two or more edges). As each node, still has
degree two, there is no base node and hence the two conditions $(M)$
and $(S)$ are again vacuously satisfied.

\item If the path has zero edge (both the coloured edges are
adjacent), let $v$ be the node common to these coloured edges. Alice
colours an edge which is incident on the node $v$ (this edge will
always exist unless sub-tree is completely coloured). Then the node $v$
will now be of degree $3$ and will become a base node.  As all the
coloured edges are incident on node $v$, so they are matched, 
hence condition $(M)$ is satisfied.  As there is only one base node
$v$ so condition $(S)$ also holds. 

\end{enumerate}

\end{enumerate}

\subsection{Case $x\geq 3$}

Next we prove the following lemma which will be used later:

{\textbf{Lemma 1}}
If condition $(S)$ is satisfied in any $x$-cl edge component then
there can be at most $\Delta$ coloured edges in that component.

Proof: In a $x$-cl edge component (say) $T$ if condition $(S)$ is true
then there can be at most one base node (say $v$) in the induced
sub-tree.  Moreover each node in any path going out of $v$ has degree
at most $2$ in the induced sub-tree. As degree of $v$ is at most
$\Delta$, there can be at most $\Delta$ paths leaving $v$. Since, each
path ends at a coloured leaf edge, there can be at most $\Delta$
coloured edges in the induced sub-tree. Q.E.D.

{\textbf{Lemma 2}}
In every $x$-cl edge component which still has at least one uncoloured
edge, if condition $(S)$ and $(M)$ are both satisfied then coloured
edges can be of at most $\Delta-1$ different colours (however the 
total number of coloured edges can be at most $\Delta$ in any
component).

Proof: Let us assume that $\gamma $ edges are incident at the base
node $v$. As condition $(M)$ holds, there can be at most $\max(3 -
\gamma, 0)$ unmatched edges in the component.  We show that if there
are more then $\Delta-1$ different colours present in the component
then condition $(M)$ can not be satisfied unless the component is
completely coloured. 

For contradiction, assume $\Delta$ or more different colours are
present in component $T$. From Lemma 1, there can not be more than
$\Delta$ coloured edges. As all edges have different colours, so there
are no matched edges except the $\gamma $ edges incident on the base
node $v$.  Thus, the number of different colours is at most $\gamma
+\max(3-\gamma,0)=\max(3,\gamma)$.  The largest possible value for
$\gamma $ is $\Delta$.  But if $\gamma =\Delta$, the component becomes
completely coloured; so we need not consider that case. We are left
with the case when $\gamma \leq \Delta-1$. Hence we can have edges of
at most $\max(3,\gamma)\leq \max(3,\Delta-1)\leq \Delta-1$ different
colours in any component (which is not fully coloured). Q.E.D.

Remark: Last inequality follows as $\Delta\geq 4$.

In rest of this section we assume that all components containing
uncoloured edges are star-like and condition $(M)$ and $(S)$ are
satisfied.  Alice chooses a component $T$ with a base node $v$, which
still has uncoloured edges.

First consider the case when all the coloured edges are incident on
the base node $v$; as a result there is no unmatched edge in the
component. Alice colours an uncoloured edge incident on $v$ (such an
edge will necessarily exist unless entire component is a star and
completely coloured). As Alice is not creating any unmatched edge so
condition $(M)$ continues to hold and as it is still star like so
condition $(S)$ also holds.

We are left with the case when there are some coloured edge (s) not
incident on our base node $v$. Let $ab$ be such an unmatched edges; in
case all coloured edges not incident at $v$ are matched, let $ab$ be
any one of them. Let $vw$ be the edge incident at $v$ which lies on
the path from node $v$ to edge $ab$, assume $a$ is closer to $v$. Path
$wa$ may contain zero or some edges. Alice colours edge $vw$ with a
feasible colour. From Lemma 2, there can be at most $\Delta-1$
different colours in the component, as we can use $\Delta+1$ colours,
there will always be a feasible colour.

After colouring edge $vw$ the tree $T$ gets split and the edge $ab$ is
not in the sub-tree (say $T'$) with base node $v$. If all coloured
edges were matched, then in $T'$ there were no unmatched edges earlier
and as we are not creating any new unmatched edge, so $(M)$ will
continue to be satisfied.  As $T'$ is star like, $(S)$ is also
satisfied. If $ab$ was unmatched, after Alice's move $\gamma $ goes up
by one, but one unmatched edge gets removed from the component, so
$(M)$ is again satisfied. Condition $(S)$ is also satisfied as $T'$ is
star like. 

Condition $(M)$ and $(S)$ also holds for the sub-tree $T''$ as it is a
path.

\subsection{Bob's move violates $(S)$} 

Next we show that

{\textbf{Lemma 3}}
In every $3$-cl edge component condition $(S)$ and $(M)$ are always
satisfied.

Proof: The induced sub-tree of $3$-cl edge component is (by
definition) cycle free and contains $3$ coloured leaf edges. As $3$
leaf edges can not be in a single path, therefore in the induced
sub-tree there must be a node of degree $3$ (say $v$). Node $v$ is a
base node (definition of base node) and there are three paths going
out of the node $v$ in the induced sub-tree. Each path is ending at a
coloured leaf edge and (in each path) every internal node is of degree
exactly $2$. Thus all the nodes other than $v$ have degree at most
$2$, so it is not possible to have another node (other than $v$) with
degree $3$ in a $3$-cl edge component. As there is exactly one base
node ($v$) in the component, so condition $(S)$ is always satisfied.

Condition $(M)$ states that every relevant $x$-cl edge star has at
most $\max(3 - \gamma, 0)$ unmatched coloured edges.  Colours of (say)
$z$ unmatched edges will be different from the $\gamma $ colours
present at the base node. Therefore the number of coloured edges is at
least $\gamma+z$. As in a $3$-cl edge component there are only $3$
coloured edges, so $z+\gamma\leq3$ or $z\leq3-\gamma \leq
\max(3-\gamma,0)$, hence condition $(M)$ holds in $3$-cl edge
component. Q.E.D.

In this section, we consider the case when Bob's move violates $(S)$.
It may or may not violate $(M)$. As $(S)$ is now violated, there are
two ``base nodes'' (nodes having degree $3$ in the induced sub-tree)
say $v$ (the earlier base node) and $w$ (the newly created base node)
in (say) $x$-cl edge component $T$. Note that before Bob's move both
conditions $(S)$ and $(M)$ were true.  After Bob's move the second
base node can be created only when Bob colours an edge, which is not
incident on the base node $v$.  Alice colours the first edge (one
incident at $v$) on the path from $v$ to $w$ with a feasible colour.
We show that this is always possible. As before Bob's move conditions
$(S)$ and $(M)$ were true, from Lemma 2 there could have been at most
$\Delta-1$ different colours present in the component. As Bob can
introduce only one new coloured edge, after his move there are at most
$\Delta$ (different) colours in the component. As we can use
$\Delta+1$ colours so, one colour is still available for Alice's move
(even when the path from $v$ to $w$ just consist of a single edge and
all the coloured edges are adjacent to edge $vw$).

After her move, component $T$ gets split and the component with base
node $v$ becomes star-like, so condition $(S)$ is satisfied.

The component containing $w$ has only three coloured edges hence, 
from Lemma 3, we know that the two conditions hold.

As we are only colouring an edge incident on the base node $v$, no new
un-matched edge is being created. $\gamma$, the number of edges
incident on $v$ does goes up by one, but two coloured edges which were
earlier in this component, are now in the new component (the component
with base node $w$) hence condition $(M)$ continues to be satisfied
(it was satisfied before Bob's move).

Remark: Only in the case when the path from $v$ to $w$ just consist of
a single edge and all the coloured edges of the component are adjacent
to either $v$ or $w$, Alice may be forced to use the $(\Delta+1)$th
colour.

\subsection{Bob's move violates $(M)$} 

We finally consider the case, when after Bob's move condition $(M)$ 
gets violated in a component (say) $T$ with base node $v$, but
condition $(S)$ holds. Thus, $T$ still has exactly one base node $v$.
After Bob's move the tree $T$ gets split and only one of the two tree
(say $T'$) can contain $v$. Let $T''$ be the other tree.

Bob could have coloured either a leaf edge of $T$ or an edge on a path
incident at $v$. In the first case, $T''$ contains one and in the
second case two coloured edges thus neither $(S)$ nor $(M)$ can be
violated in $T''$.

Hence, the tree $T'$ is the sub-tree which does not satisfy $(M)$. As
condition $(M)$ is being violated, there is at least one unmatched
edge, say $ab$. Alice colours the first edge on path from $v$ to $ab$
with a feasible colour.

As conditions $(S)$ and $(M)$ were true before Bob's move, so from
Lemma 2, at most $\Delta-1$ different colours were present in $T$. As
Bob can introduce only one new colour, so now at most $\Delta$ colours
are present. As Alice has $\Delta+1$ colours for colouring, there is a
feasible colour left.

Remark: In the case when $ab$ is the only unmatched edge and the path
is null, Alice may have to use the $(\Delta+1)$th colour. In all
other cases she can colour the edge with one of the $\Delta$ colours
already present.

\section{Implementation Issues}

With out loss of generality, assume that the initial forest consists
of a single tree (else, we can run the algorithm independently on each
component). 

We will keep the list of coloured leaves (and pointers to them) for
each component (after edges get split). The maximum number of coloured
leaves in a component is $\Delta$.

Given two coloured nodes there is a unique path in the original tree.
To determine the path efficiently, we root the uncoloured tree at any
vertex and preprocess the tree for lowest common ancestor (LCA)
queries. This takes $O(n)$ time, and after preprocessing, LCA queries
can be answered in $O(1)$ time [15],[13],[11]. The path between two
vertices (say) $u$ and $v$ in the tree will consist of path from $u$
to $w=$LCA$(u,v)$ (the lowest common ancestor of $u$ and $v$) and from
$w$ to $v$.

Let $u,v,w$ be three (coloured leaf) nodes in the tree. Then the
``induced'' sub-tree for these three nodes will only contain edges
between $u--v,u--w,v--w$. Let $x_{uv}=$LCA$(u,v)$,
$x_{uw}=$LCA$(u,w)$, and $x_{vw}=$LCA$(v,w)$. 

Without loss of generality, assume that $x_{uv}$ is closer to $u$ than
than $x_{uw}$. As $x_{uv}$ and $x_{uw}$ are both ancestors of $u$, the
path from $u$ to $w$ will pass through $x_{uw}$. Or $x_{uv}$ will on
both $u-v$ and $u-w$ paths. Hence, it will have degree three and will
be a base node.

Thus, to determine the base node, we just need the height (or depth)
of the nodes (in the original tree). 

In general, after Bob's move there can be at most $\Delta$ coloured
leaves edges in any component. In case, two or more leaf edges are
incident on any node, then that node is a base node. Else, we pick up
any leaf edge say $(\bullet,u)$ incident at $u$. And find four LCAs.
The one closest to $u$ will be a base node. Next base node, will be
the ancestor closest to this LCA. 

If an edge on the induced sub-graph gets coloured, the component gets
split. The coloured edges which are no longer in the earlier component
(these can be determined by LCA queries) are deleted from the list of
coloured leaves and the new coloured edge is added. The other coloured
leaves (along with the newly coloured edge) will be added to the list
of the new component.

Thus, Alice can select the edge which she should colour in $O(1)$
time, provided she knows the component in which the edge Bob just
coloured lies. Let $ab$ be the edge last coloured. If there is another
coloured edge incident at $a$ (respectively $b$), the Alice colours
another edge incident at $a$ (resp. $b$). Else, let $v$ be the base
node in the component containing $ab$--- actually, the edge $ab$ will
lie in two components, but because of condition $(S)$, only one of
them will contain a base node. Alice selects an edge incident at $v$
on the path from $v$ to $ab$.

Remark: Other condition $(M)$ ensures that even after Bob's move there
are at most $\Delta$ different colours in a component (see Lemma 2),
thus Alice will always have a feasible colour.

We next discuss the problem of maintaining components. There are
several techniques for example, ET-Trees [14], linking and cutting
trees [16] which can be used to solve the problem in $O(\log n)$ time;
in fact any method for dynamically maintaining connected components
(actually, only decremental) can be used here.  We next describe a
simple method to solve this problem.

First the basic method of Even and Shiloach[7]. To begin with the
entire tree is in a single component. Let us store the name of the
component (say $1$) with each node in the tree. When an edge (say)
$uv$ is removed, the tree splits into two parts, one containing $u$
and the other $v$. We determine the tree containing fewer node. This
can be done as in [14],[7], by start traversing both trees, in
parallel, and stopping as soon as one tree is completely traversed.
The name of component, stored in vertices of the smaller tree (one
completely traversed) is changed (to a new name). As the size of the
tree in which labels are changed is at most half the size of the
original tree, label of any node can be changed at most $\log_2 n$
times. Thus, this method takes $O(n\log n)$ time.

We next reduce the time, by having a two level structure.  First
assume that no internal node has exactly one child. We replace
sub-trees rooted at the ``lowest'' nodes having at least $\frac12 \log
n$ descendants (i.e., each child of these nodes has fewer than
$\frac12 \log n$ descendants) by a super-node. This is done until
$O(n/\log n)$ nodes are left. 

The name of component for any node inside the super-node will be the
name of the component of the super-node.

Thus, the resulting trees (with super-nodes) will have $O(n/\log n)$
nodes; if we use the basic method for this tree the time will be
$O(n)$. If an edge is deleted inside a super-node, sub-tree inside a
super-node is split. But as ``internal node'' in the sub-tree of
super-node has $O(\log n)$ nodes, the basic method will take
$O(\log\log n)$ time, per node. Thus, we are able to reduce the time
to $O(n\log\log n)$.
 
If we repeat the two-level method, on each sub-tree the time will
reduce to $O(n\log^{(3)}n)$. Similarly by using $k$-level structure we
get an $O(n\log^{(k)}n)$ time method. 

Each two-level structure can also be preprocessed using the method of
``micro-sets'' [12] to get a linear time algorithm or we can use a
method similar to that of[10] to get a near linear time algorithm
(which may have smaller constants). Basically, instead of using
logarithms, we can use Ackermann function and get a bound of $O(n
\alpha(n))$, here $\alpha( )$ is inverse Ackermann function.  

Finally, assume that there are nodes with exactly one child. We
replace this ``chain'' by a single edge to get ``compressed tree''. In
compressed tree each node has at least two children and we preprocess
it as before. Maintaining linear chain of descendants is equivalent to
the set splitting problem studied by Gabow and Tarjan[10][12] or list
splitting problem studied by Gabow[10] and can be solved in above
bounds.

\section*{References}

\begin{enumerate}
\item Andres, D. (2003). The game chromatic index of forests of
maximum degree 5, Electronic Notes in Discrete Mathematics,
13:5--8.

\item Andres, S. D. (2006). The game chromatic index of forests of
maximum degree $\delta\geq5$. Discrete Applied Mathematics,
154(9):1317--1323.

\item Bodlaender, H. L. (1990). On the complexity of some coloring
games. In International Workshop on Graph-Theoretic Concepts in
Computer Science, pages 30--40. Springer.

\item Cai, L. and Zhu, X. (2001). Game chromatic index of
k-degenerate graphs. Journal of Graph Theory, 36(3):144--155.

\item Chan, W. H. and Nong, G. (2014). The game chromatic index of
some trees of maximum degree 4. Discrete Applied Mathematics,
170:1--6.

\item Erdos, P. L., Faigle, U., Hochstattler, W., and Kern, W.
(2004). Note on the game chromatic index of trees. Theoretical
Computer Science, 313(3):371--376.

\item Even, S. and Shiloach, Y. (1981). An on-line edge-deletion
problem. Journal of the Association for Computing Machinery,
28(1):1--4.

\item Fong, W. L., Chan, W. H., and Nong, G. (2018). The game chromatic
index of some trees with maximum degree four and adjacent degree-four
vertices. Journal of Combinatorial Optimization, 36:1--12.

\item Fong, W. L., and Chan, W. H. (2019). The edge coloring game on
trees with number of colors greater than the game chromatic index.
Journal of Combinatorial Optimization, 38:456--480.

\item Gabow, H.N., (1985), A scaling algorithm for weighted matching
on general graphs Prof. 26th IEEE FOCS, 90--100.

\item Gabow, H. N., Bentley, J. L., and Tarjan, R.E., (1984). Scaling and
related techniques for geometry problems, STOC '84: Proc. 16th ACM
Symposium on Theory of Computing, New York, NY, USA: ACM, 135--143

\item Gabow, H.N., and Tarjan, R.E. (1985). A linear-time algorithm
for a special case of disjoint set union. Journal of Computer and
System Science, 30:209-221.

\item
Harel, D., and Tarjan, R.E. (1984).
Fast algorithms for finding nearest common ancestors.
SIAM Journal on Computing, 13(2): 338--355,

\item Henzinger, M. R. and King, V. (1995).  Randomized dynamic graph
algorithms with polylogarithmic time per operation.  Proceedings of
the twenty-seventh annual ACM symposium on Theory of computing - STOC
'95, 519-527

\item Schieber, B., and Vishkin, U. (1988). On finding lowest common
ancestors: simplification and parallelization. SIAM Journal on
Computing, 17(6): 1253--1262

\item Sleator, D. D., and Tarjan, R. E. (1985).  Self-Adjusting
Binary Search Trees. Journal of the ACM., 32(3): 652--686.
\end{enumerate}
\end{document}